\documentclass[preprint,superscriptaddress,showkeys,tightenlines,showpacs]{revtex4} 
\usepackage{graphicx}
\begin{document}
\preprint{PNU-NTG-07/2006}
\preprint{PNU-NURI-08/2006}
\title{Test of the reaction mechanism for $\gamma N\to K\Lambda(1520)$ 
  using the polarized photon}
\author{Seung-il Nam}
\email{sinam@pusan.ac.kr}
\affiliation{Department of Physics and Nuclear physics \& Radiation
Technology Institute (NuRI), Pusan National University, Busan 609-735,
Korea}  
\author{Ki-Seok Choi}
\email{kschoi@pusan.ac.kr}
\affiliation{Department of
Physics and Nuclear physics \& Radiation Technology Institute (NuRI),
Pusan National University, Busan 609-735, Korea} 
\author{Atsushi Hosaka}
\email{hosaka@rcnp.osaka-u.ac.jp}
\affiliation{Research Center for Nuclear Physics (RCNP), Ibaraki, Osaka
567-0047, Japan}
\author{Hyun-Chul Kim}
\email{hchkim@pusan.ac.kr}
\affiliation{Department of
Physics and Nuclear physics \& Radiation Technology Institute (NuRI),
Pusan National University, Busan 609-735, Korea} 
\date{November 2006}
\begin{abstract}
We study the reaction mechanism for the  photoproduction of 
the $\Lambda(1520)$, based on an effective Lagrangian approach.  We
investigate each contribution of the $s$-, $u$-, $t$-channel processes 
and contact term, separately.  One of the most characteristic features
of this reaction is the contact-term dominance which governs the
photoproduction from the proton, when the $K^*$-exchange contribution
is possibly not too large.  We suggest 
several different ways of the polarizations and arrangement 
of the beam and target to make to understand the role of each
contribution separately in future experiments.
\end{abstract}
\maketitle
\section{Introduction}
Recent activities in hadron physics are largely motivated by 
the observation signals for exotic hadrons.  
Perhaps, one of the most influential ones is the $\Theta^+$.  
After the great amount of efforts, we are now at the stage of 
more careful investigation on the matter.  
The study of the exotic hadrons, at the same time, has brought
revised interest in the physics of hadron resonances in general.  
Among them, particularly interested is the sector of 
strangeness $S=-1$.  The properties of the $\Lambda(1405)$ as well as
the $\bar K$-$N$ interaction have been discussed in many different
ways: One noticeable suggestion was made for the $\Lambda(1405)$ by 
regarding it as the two pole structure of the resonance.   

Another resonance $\Lambda(1520) \equiv \Lambda^*$ attracts also
a great deal of attention.  This resonance of $J^P = 3/2^-$ is well
established with a relatively narrow decay width $\Gamma \sim 15$ MeV.   
In the quark model it is a member of 70-plet of SU(6) with $l = 1$
excitation, and is mostly dominated by the flavor singlet component.   
In the dynamical treatment of meson-baryon scattering, 
it appears as a quasi-bound state of $\pi$ and $\Sigma(1385)$.  
Both methods are  equally successful in explaining the currently 
existing data of the $\Lambda^*$.  However, they have very different
structure in its wave function, and make different predictions in some
physical quantities.   

In modern experimental facilities such as the SPring-8 and Jefferson
Laboratory (JLab), production experiments of this resonance can be carried
out with high precision, so that it is of great interest and
importance to study the reactions theoretically.  Understanding the
mechanism of the $\Lambda^*$-production is very useful not only for 
the extraction of its properties but also for the explanation of the 
recent observation of the peak structure corresponding to the
$\Theta^+$ in the reaction $\gamma +d \to \Theta^+ + \Lambda^*$.  
Motivated by these facts, there have been several works on this 
reaction both experimentally~\cite{Nakano:2005cg} and
theoretically~\cite{Roca:2004wt,Nam:2005uq,Nam:2005yg,Titov:2005kf},
very recently.

In addition to the previous experiments for the
$\Lambda^*$-photoproduction~\cite{Boyarski:1970yc,Barber:1980zv}, 
a recent experiment conducted by the LEPS collaboration at the
SPring-8 revealed an interesting feature: It turned out that the
production rate of the $\Lambda^*$ was strongly dependent on the target  
nucleon, which implies that the production mechanism of the
$\Lambda^*$-photoproduction has strong charge asymmetry.  
Interestingly, this feature is consistent with the theoretical
prediction of Ref.~\cite{Nam:2005uq} in which it was proposed that  
this asymmetry might be explained by the contact term which was 
required in order to satisfy the gauge invariance, and was 
present for the proton target ($\gamma p \to K^+ \Lambda^*$) but not  
for the neutron target ($\gamma n \to K^0 \Lambda^*$).  
The size of the contact term is significant, at least 
as compared with the $s$-channel, $u$-channel, and kaon exchange in 
the $t$-channel process.  The contact-term dominance shown in the
previous study~\cite{Nam:2005uq}, however, depends on the unknown
strength of the $K^*$-exchange diagram which is active both for the
proton and neutron in the almost same order of magnitude.  
The importance of the contact term was ignored in the analysis 
of the previous experiments~\cite{Boyarski:1970yc,Barber:1980zv},
while the role of the $K^*$-exchange was emphasized.

It is now of great importance to test each contribution of various
terms in the reaction, especially, that of the contact and
$K^*$-exchange terms. In the present work, we aim at investigating
this issue in very detail.  In the current and future experiments such
as the LEPS and the CLAS at the JLAB, various angular distributions
are expected to be available with the polarized photon and nucleon
target utilized.  Therefore, we consider the differential cross
sections under different conditions of the polarization of the beam
and target so that we may extract information on how each term
contributes to the differential cross sections.  

This paper is organized as follows: In Section 2, we explain the
present method briefly.  We define various effective interactions
which are necessary to compute the transition amplitudes.  In Section
3, we present the differential cross sections with each Born diagram
separately,  including the $s$-, $u$-, $t$-channel processes and
contact term.  In Section 4, we consider possible experimental
conditions to distinguish the contact term dominance from $K$- and
$K^*$-exchanges by using angular distributions and asymmetries.  The final 
Section is devoted to summary and conclusions.     

\section{Effective Lagrangian method}
\begin{figure}[t]
\includegraphics{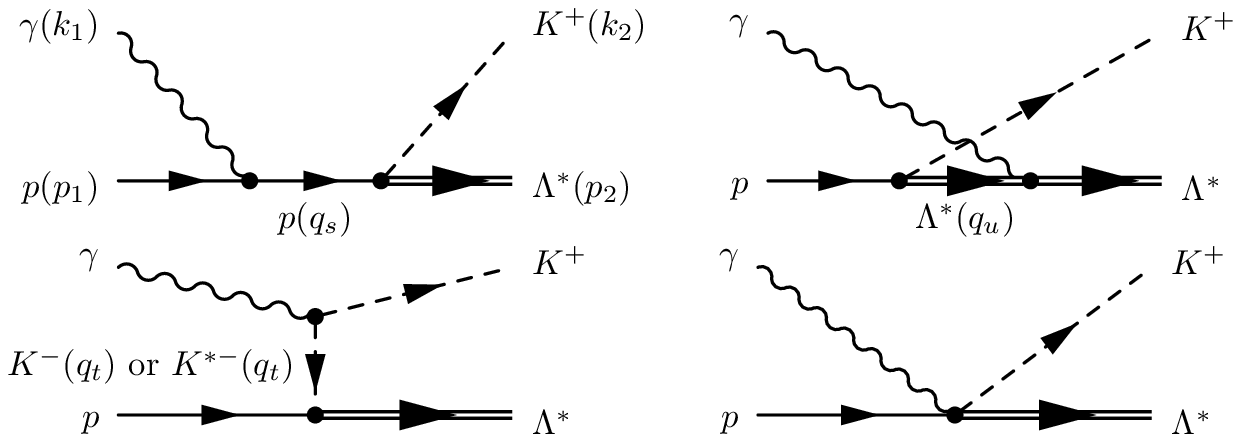}
\caption{The relevant diagrams for the $\Lambda^*$-photoproduction.}
\label{fig0}
\end{figure}

In this Section we define the effective Lagrangians relevant to the
$\gamma p\to K^+\Lambda^{*}$ process as depicted in Fig.~\ref{fig0},  
where we consider five tree (Born) diagrams: $s$-channel (upper-left),
$u$-channel (upper-right), $t$-channel with pseudoscalar
$K^-$-exchange ($t(p)$) and with vector $K^{*-}$-exchange ($t(v)$)
(lower-left), and contact term (lower-right). 
We define the momenta of the photon, pseudoscalar kaon, vector kaon,
nucleon and $\Lambda^*$ as shown in Fig.~\ref{fig0}.
The spin-3/2 particle is treated as a Rarita-Schwinger 
(RS) field~\cite{Rarita:mf,Nath:wp,Hagen:ea}, and its 
treatment can be found in Ref.~\cite{Nam:2005uq} in detail. 
The relevant effective Lagrangians are then given as follows:
\begin{eqnarray}
\mathcal{L}_{\gamma NN}&=&-e\bar{N}\left(\rlap{/}{A}
+\frac{\kappa_{N}}{4M_{p}}\sigma_{\mu\nu}F^{\mu\nu}\right)N\,
+{\rm h.c.},\nonumber\\
\mathcal{L}_{\gamma KK}&=&
ie\left\{ 
(\partial^{\mu}K^{\dagger})K-(\partial^{\mu}K)K^{\dagger}
\right\}A_{\mu},\nonumber\\
\mathcal{L}_{\gamma
\Lambda^*\Lambda^*}&=&-e\bar{\Lambda^*}^{\mu}\left(\rlap{/}{A}+
\frac{\kappa_{\Lambda^*}}{4M_{\Lambda^*}}\sigma_{\nu\rho}
F^{\nu\rho}\right)\Lambda^*_{\mu}\,+{\rm 
h.c.},\nonumber\\
\mathcal{L}_{\gamma
  KK^{*}}&=&g_{\gamma
  KK^{*}}\epsilon_{\mu\nu\sigma\rho}(\partial^{\mu}A^{\nu})
(\partial^{\sigma}K)K^{*\rho}\,+{\rm 
h.c.},\nonumber\\
\mathcal{L}_{KN\Lambda^*}&=&\frac{g_{KN\Lambda^*}}{M_{K}}
\bar{\Lambda}^{*\mu}
\Theta_{\mu\nu}(A,Z)(\partial^{\nu}K){\gamma}_{5}p\,+{\rm 
h.c.},\nonumber\\
\mathcal{L}_{K^{*}N\Lambda^*}&=&-\frac{ig_{K^{*}N\Lambda^*}}
{M_{K^*}}\bar{\Lambda}^{*\mu}\gamma^{\nu}(\partial_{\mu}
K^{*}_{\nu}-\partial_{\nu}K^{*}_{\mu})p+{\rm 
h.c.},\nonumber\\ \mathcal{L}_{\gamma
KN\Lambda^*}&=&-i\frac{eg_{KN\Lambda^*}}{M_{K}}
\bar{\Lambda}^{*\mu}A_{\mu}K{\gamma}_{5}N\,+{\rm
h.c.},
\label{Lagrangian}
\end{eqnarray}
where $N$, $\Lambda^*_{\mu}$, $K$, $K^*$ and $A^{\mu}$ are the
nucleon, $\Lambda^*$, pseudoscalar kaon, vector $K^*$ and photon  
fields, respectively.  The interaction for the $K^{*}N\Lambda^*$
vertex is taken from Ref.~\cite{Machleidt:1987hj}.  As for the $\gamma
\Lambda^* \Lambda^*$ vertex for the $u$-channel, we utilize the
effective interaction suggested by Ref.~\cite{gourdin} which contains
four form factors of different multipoles. We ignore the electric
coupling $F_1$, since the $\Lambda^*$ is neutral. We also neglect
$F_3$ and $F_4$ terms, assuming that higher multipole terms are less
important. Hence, for the photon coupling to the $\Lambda^*$, we
consider only the magnetic coupling term $F_2$ whose strength is
proportional to the anomalous magnetic moment of the $\Lambda^*$,
$\kappa_{\Lambda^*}$.  In the present work, we set
$\kappa_{\Lambda^*}=1$ as a trial.  We will see that the $u$-channel
contribution is very small as compared with the dominant contact term
if the $\kappa_{\Lambda^*}$ is not too large.  The $KN\Lambda^*$
coupling constant, $g_{KN\Lambda^*}$ is determined to be $11$ by using 
the experimental data ($\Gamma_{\Lambda^*\to\bar{K}N}=15.6\times 45\%$ 
MeV)~\cite{Yao:2006px}.  We will discuss the $K^*N\Lambda^*$ coupling
constant shortly.  We note that the last Lagrangian in
Eq.~(\ref{Lagrangian}) represents the contact term interaction which is
necessary to maintain the Ward-Takahashi identity of the amplitudes. 

Having set up these Lagrangians, we can write the scattering
amplitudes as follows:
\begin{eqnarray}
i\mathcal{M}_{s}&=&-\frac{eg_{KN\Lambda^*}}{M_{K}}\bar{u}^{\mu}(p_{2},s_{2})
k_{2\mu}{\gamma}_{5} \frac{(\rlap{/}{p}_{1}+M_{p})F_{c}+
\rlap{/}{k}_{1}F_{s}}{q^{2}_{s}-M^{2}_{p}}
\rlap{/}{\epsilon}u(p_{1},s_{1}),\nonumber\\&+&
\frac{e\kappa_{p}g_{KN\Lambda^*}}{2M_{p}M_{K}}
\bar{u}^{\mu}(p_{2},s_{2})k_{2\mu}{\gamma}_{5} 
\frac{(\rlap{/}{q}_{s}+M_{p})F_{s}}{q^{2}_{s}-M^{2}_{p}}
\rlap{/}{\epsilon}\rlap{/}{k}_{1}u(p_{1},s_{1})\nonumber\\
i\mathcal{M}_{u}&=&-\frac{g_{KN\lambda}\kappa_{\Lambda^*}}{2M_{K}M_{\Lambda}}
\bar{u}_{\mu}(p_2)\rlap{/}{k}_{1}\rlap{/}{\epsilon}D^{\mu}_{\sigma}
\Theta^{\sigma\rho}k_{2\rho}\gamma_{5}u(p_1)F_{u},\nonumber\\ 
\mathcal{M}_{t}&=&\frac{2eg_{KN\Lambda^*}}{M_K}
\bar{u}^{\mu}(p_{2},s_{2})\frac{q_{t,\mu}k_{2}
\cdot\epsilon}{q^{2}_{t}-M^{2}_{K}}
{\gamma}_{5}u(p_{1},s_{1})F_{c},\nonumber\\
i\mathcal{M}_{c}&=&\frac{eg_{KN\Lambda^*}}{M_K}\bar{u}^{\mu}(p_{2},s_{2})
\epsilon_{\mu}{\gamma}_{5}u(p_{1},s_{1})F_{c},\nonumber\\i\mathcal{M}_{v}&=&
\frac{-ig_{\gamma{K}K^*}g_{K^{*}NB}}{M_{K^{*}}(q^{2}_{t}-M^{2}_{K^*})}   
\bar{u}^{\mu}(p_{2},s_{2})\gamma_{\nu}\left(q^{\mu}_{t}g^{\nu\sigma}-
g^{\nu}_{t}q^{\mu\sigma}\right)\epsilon_{\rho\eta\xi\sigma}k^{\rho}_{1}
\epsilon^{\eta}k^{\xi}_{2}u(p_{1},s_{1})F_{v}\, , 
\label{amplitudes}  
\end{eqnarray}
where $\epsilon$ and ${u}^{\mu}$ are the photon polarization vector
and the RS vector-spinor.  $D_{\mu\nu}$ is the spin-3/2 fermion 
propagator~\cite{Read:ye,Nam:2005uq}, for which here we use the usual
spin-1/2 fermion propagator, since the difference is negligible 
in the relatively low energy regions~\cite{Nam:2005uq}.  For the form 
factor, we employ the gauge invariant one with the four momentum 
cutoff $\Lambda$, which was used for the study of $\gamma N\to
K\Lambda^*$ in Ref.~\cite{Nam:2005uq}: 
\begin{eqnarray}
F_{x}(q^2)&=&\frac{\Lambda^4}{\Lambda^4+(x-M^2_x)^2},\,\,x=s,u,t(p),t(v),
\nonumber\\
F_{c}&=&F_s+F_u-F_sF_u \, .
\label{formfactor}
\end{eqnarray}    
Here $M_x$ is the on-shell mass of the exchange particle in the 
$x$-channel.  In Ref.~\cite{Nam:2005uq}, it was shown that the use of
this form factor in Eq.(\ref{formfactor}) satisfies the Ward-Takahashi
identity and can reproduce the existing data qualitatively well.  

In the present treatment there are two unknown parameters, i.e. 
the cutoff $\Lambda$ in the form factor and the $K^* N\Lambda^*$
coupling constant.  The former is fixed to be $\Lambda=700$ MeV as 
done in Refs~\cite{Nam:2004fh,Nam:2005uq}.  The only remaining unknown 
is the strength of the $K^* N\Lambda^*$ coupling constant.  In
theoretical models, the coupling constant is rather
small~\cite{Hyodo:2006uw}.  Since this value affects the feature of
the contact term dominance, we will treat it as a free parameter and
see how much the contact term contributes to the amplitude.   

\section{Results and Discussion}
\subsection{Dominant processes}
In Ref.~\cite{Nam:2005uq,Nam:2005yg}, it was already shown that the 
contribution of the contact term was significantly larger than those
from the $s$-, $u$-, $t$-channels and $K^*$-exchange when 
$|g_{K^*N\Lambda^*}|$ was not too large, i.e. $|g_{K^*N\Lambda^*}|
\lesssim 10$.  In order to verify it, we need to examine each
contribution to the differential cross sections in detail.   

\begin{figure}
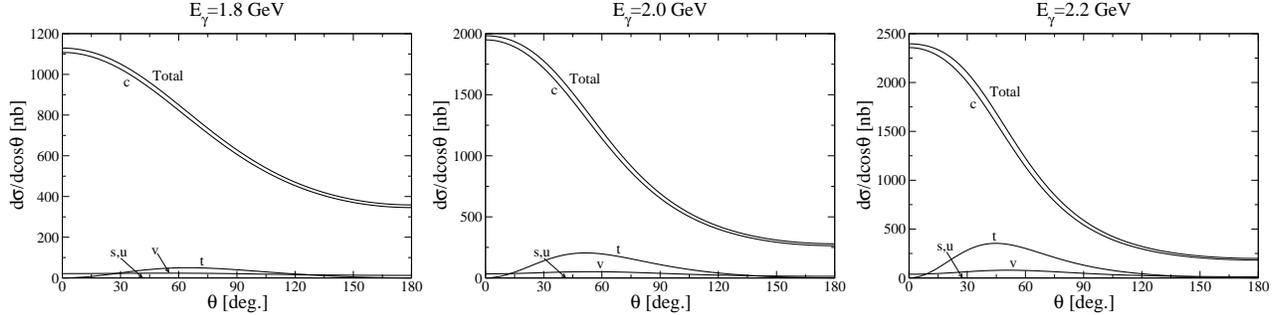

\centerline{
\includegraphics[width=5.5cm]{f1.eps}
\includegraphics[width=5.5cm]{f2.eps}
\includegraphics[width=5.5cm]{f3.eps}}
\caption{Each contribution to the differential cross sections as a
function of $\theta$ at three different photon energies
$E_{\gamma}=1.8$, $2.0$ and $2.2$ GeV and the differential cross
section with all contributions.  The c, s, u, and v denote the
contact term, the $s$-channel, the $u$-channel, and $K^*$-exchange,
respectively.}   
   \label{theta_dep}
 \end{figure}

In Fig.~\ref{theta_dep}, we draw the differential cross sections for
each contribution separately as well as that with all contributions as 
functions of the scattering angle $\theta$ of the kaon at three
different photon energies, $E_{\gamma}=1.8$, $2.0$ and $2.2$ GeV,
respectively.  The $K^*N\Lambda^*$ coupling constant is 
fixed to be $g_{K^*N\Lambda^*} = 11$ for $K$-exchange.  We observe
from Fig.~\ref{theta_dep} that the differential cross section
with all contributions increases in particular in the forward
direction as $E_\gamma$ increases.  As shown in  Fig.~\ref{theta_dep},
the contact term is the most dominant one that makes almost all
contributions to the differential cross section, even though the sizes
of the $K$- and $K^*$-exchange contributions are still non-negligible.
On the contrary, those of the $s$- and $u$-channels turn out to be
tiny.  Therefore, it is sufficient to consider the contact term, $K$-
and $K^*$-exchanges.    

\subsection{$\phi$-dependence with polarized photon beams}
\begin{figure}
\centerline{
\includegraphics[width=7cm]{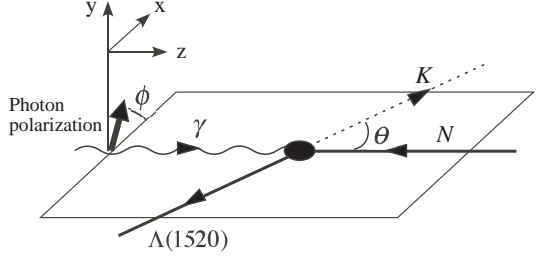}}
\caption{Definition of axes on the reaction plane and scattering  
and azimuthal angles $\theta$ and $\phi$.} 
\label{def_axis}
 \end{figure}

It is of great interest to determine which particle is exchanged in
the course of the photoproduction.  In fact, it is possible to
pinpoint the exchange particle if the photon beam can be polarized in
various ways.  In this case, the dependence on the azimuthal angle
($\phi$) becomes important in addition to the scattering angle.  In
order to study the dependence of each channel on the $\phi$, we take
the $z$-axis for the incident photon beam, the $x$-axis for the
scattering plane, and the $y$-axis to be perpendicular to the
scattering plane (see Fig.~\ref{def_axis}).   If we take a linearly  
polarized photon, then the $\phi$ can be defined as that between the
photon polarization and the $x$-axis. 

\begin{figure}
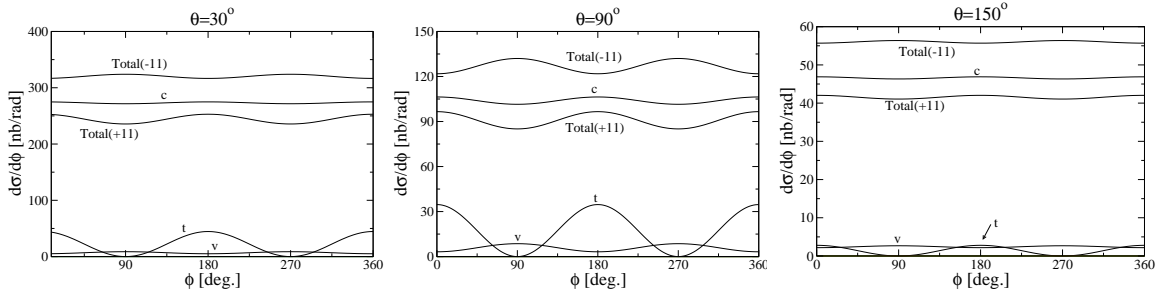

\centerline{
\includegraphics[width=5cm]{f5.eps}
\includegraphics[width=5cm]{f6.eps}
\includegraphics[width=5cm]{f7.eps}}
\caption{Each contribution to the differential cross sections as a
function of $\phi$ at three different scattering angles 
$\theta = 30^\circ$, $90^\circ$ and $150^\circ$.
The photon energy is fixed to be $E_\gamma = 2$ GeV, and 
$g_{K^*N\Lambda^*}=\pm11$.  The c, t, and v denote the
contact term, the $t$-channel, and $K^*$-exchange, respectively.}
\label{phi_dep}
 \end{figure}

Let us first show how different $\phi$-dependence emerges according to  
the type of the electromagnetic interaction.  This is particularly
useful to distinguish one type from the other in meson exchange
diagrams of the $t$-channel.  Consider the $K$-exchange amplitude
which contains the $\gamma KK$ coupling constant.  The coupling
constant is an electric type and has the following structure:  
$\epsilon \cdot (\vec q + \vec p_f)$, where $\vec q$ and $\vec p_f$
are the momenta of the exchanged (intermediate) and final-state kaons.
Therefore, the scattered kaon and nucleon in the final state are
produced most likely to be in the direction parallel to the photon
polarization vector.  Explicitly, the resulting $\phi$-dependence of
the cross section is proportional to $\cos^2 \phi$, as shown in
Fig.~\ref{phi_dep}.  For $K^*$-exchange, there are several terms for
the $\gamma KK^*$, but the important piece has a structure of the
magnetic type, i.e. $\vec \epsilon \times \vec q \cdot \vec p_f$.
Hence, the final-state kaon and nucleon are produced most likely in
the direction perpendicular to the photon polarization, which is
proportional to $\sin^2 \phi$.  This behavior can be also verified as
shown in Fig.~\ref{phi_dep}, although it is not exactly $\sin^2 \phi$.   
Different angular distributions depending on which particle is 
exchanged was used in the analysis of the $\phi$-photoproduction and
in the study of $\Lambda(1405)$-photoproduction.  As compared to the
$t$-channel processes, the $\phi$-dependence is not so strong for the 
contact term.  

As an alternative method instead of the $\phi$-dependence, it is also
convenient to consider the asymmetry defined as  
\begin{eqnarray}
A(E_{\gamma},\theta)=\frac{\left(\frac{d\sigma}{d\theta}\right)_{\perp}
-\left(\frac{d\sigma}{d\theta}\right)_{\parallel}}{
\left(\frac{d\sigma}{d\theta}\right)_{\perp}
+\left(\frac{d\sigma}{d\theta}\right)_{\parallel}}.
\label{BADCS}
\end{eqnarray} 
The subscripts $\parallel$ and $\perp$ in Eq.~(\ref{BADCS}) stand for 
the direction of the photon polarization vectors parallel and
perpendicular to the reaction plane, respectively. 
The asymmetry integrated over the scattering angle is then defined as:  
\begin{eqnarray}
A(E_{\gamma})=\frac{1}{2}\int A(E_{\gamma},\theta)\sin\theta d\theta.
\label{BATCS}
\end{eqnarray}  
Coefficient $1/2$ on the r.h.s. is introduced for the normalization of  
$A(E_{\gamma})$ such that it reduces to either $1$ or $-1$ when
$(d\sigma / d\theta)_{\perp}$ or $(d\sigma / d\theta)_{\parallel}$
vanishes.    

\begin{figure}[t]
\begin{tabular}{c}
\includegraphics[width=8cm]{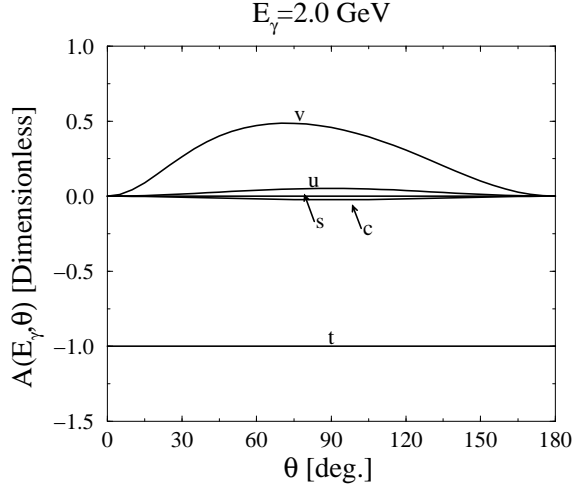}
\end{tabular}
\caption{Asymmetry $A(E_{\gamma},\theta)$ for various contributions.}    
\label{phi_dep1}
\end{figure}

\begin{figure}[t]
\begin{tabular}{cc}
\includegraphics[width=6cm]{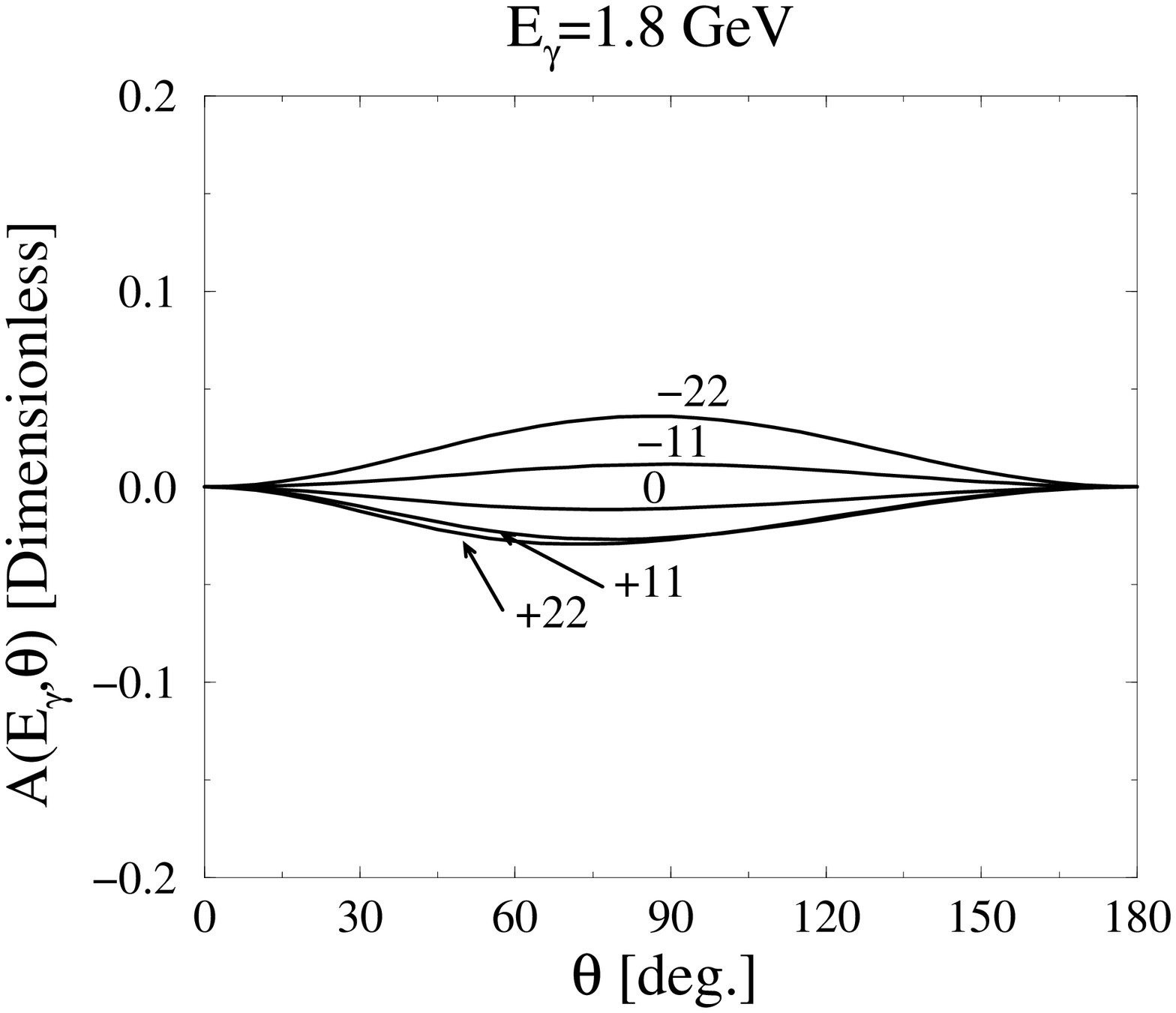}
\includegraphics[width=6cm]{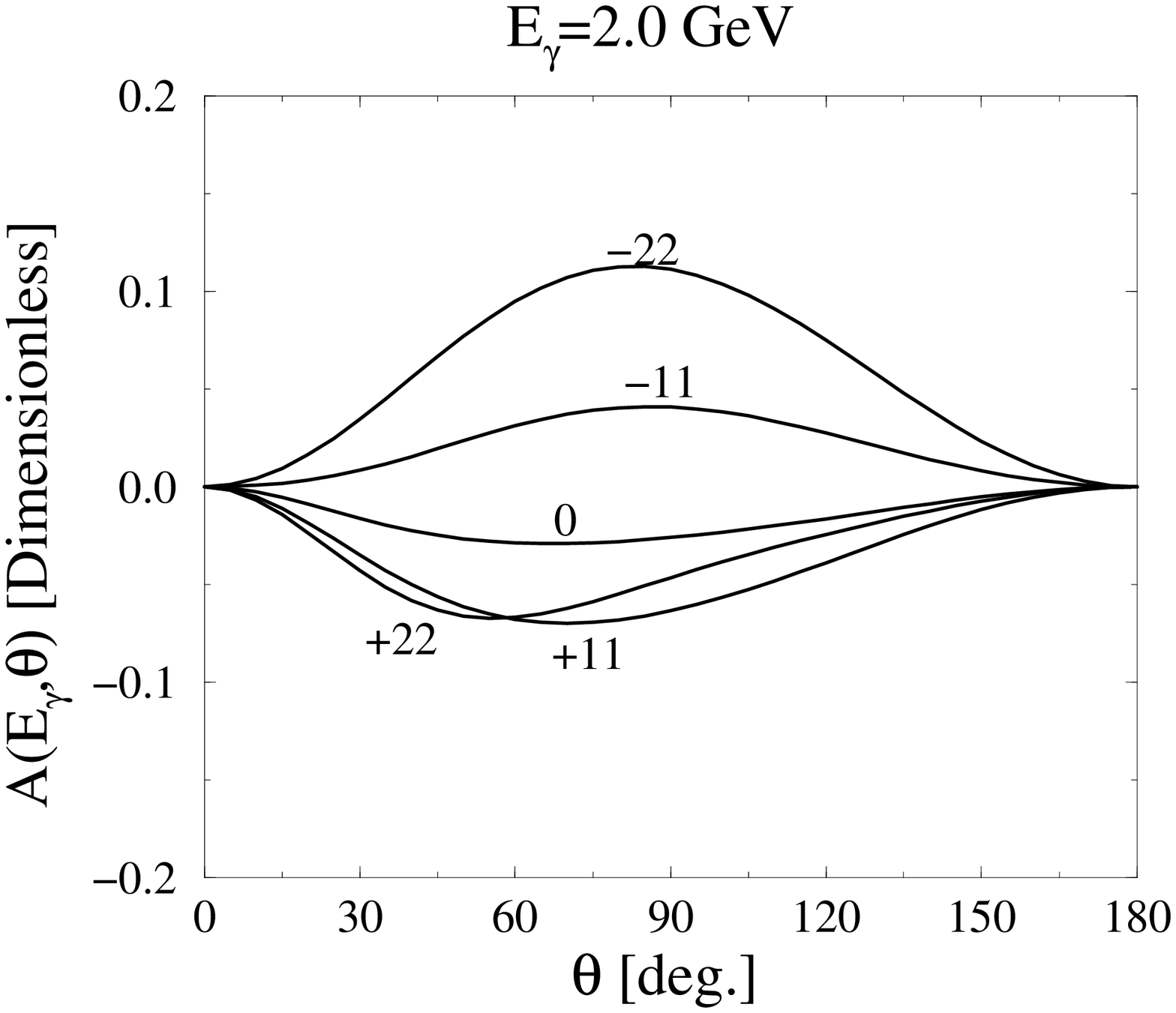}
\end{tabular}
\begin{tabular}{c}
\includegraphics[width=6cm]{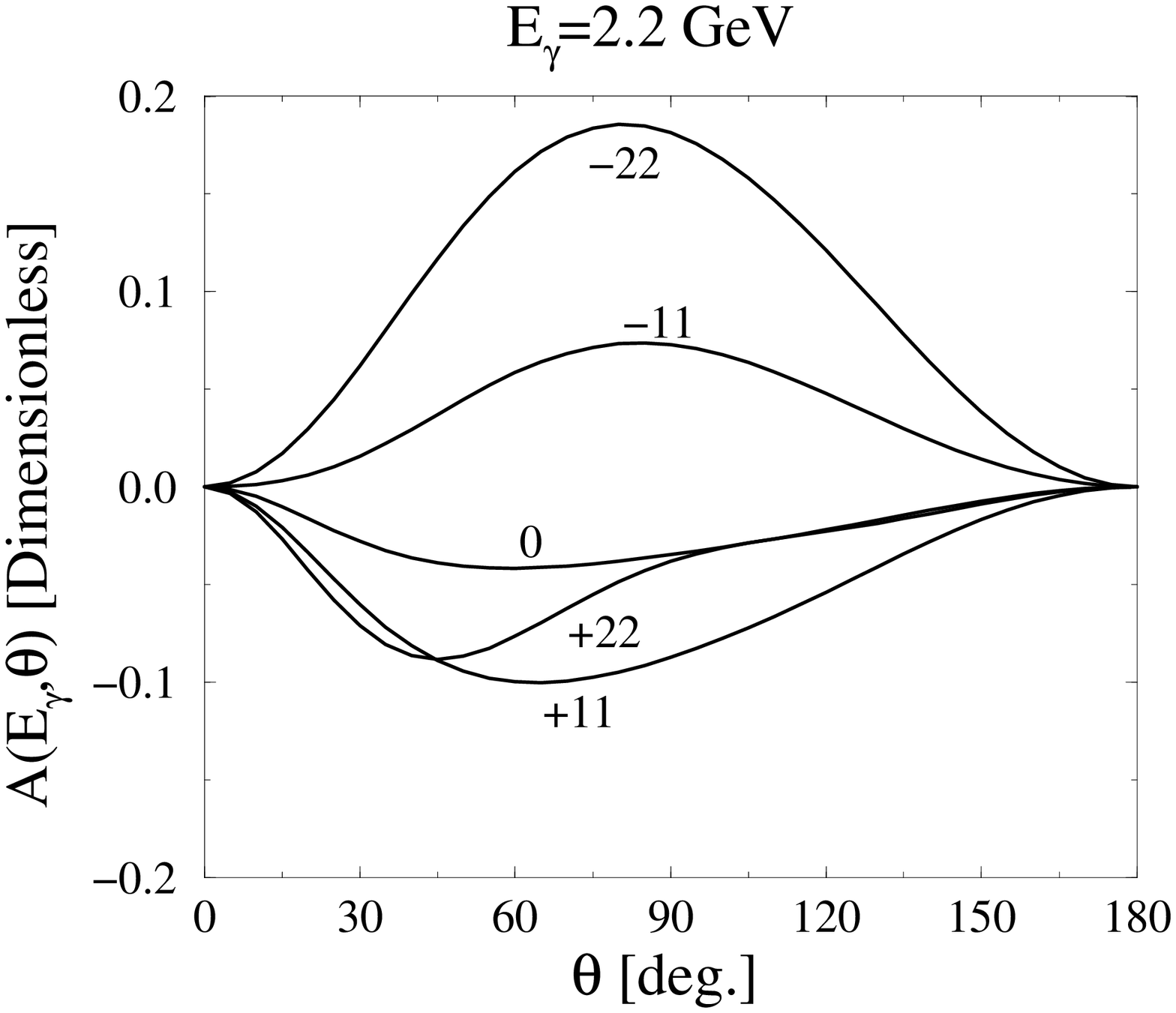}
\includegraphics[width=6cm]{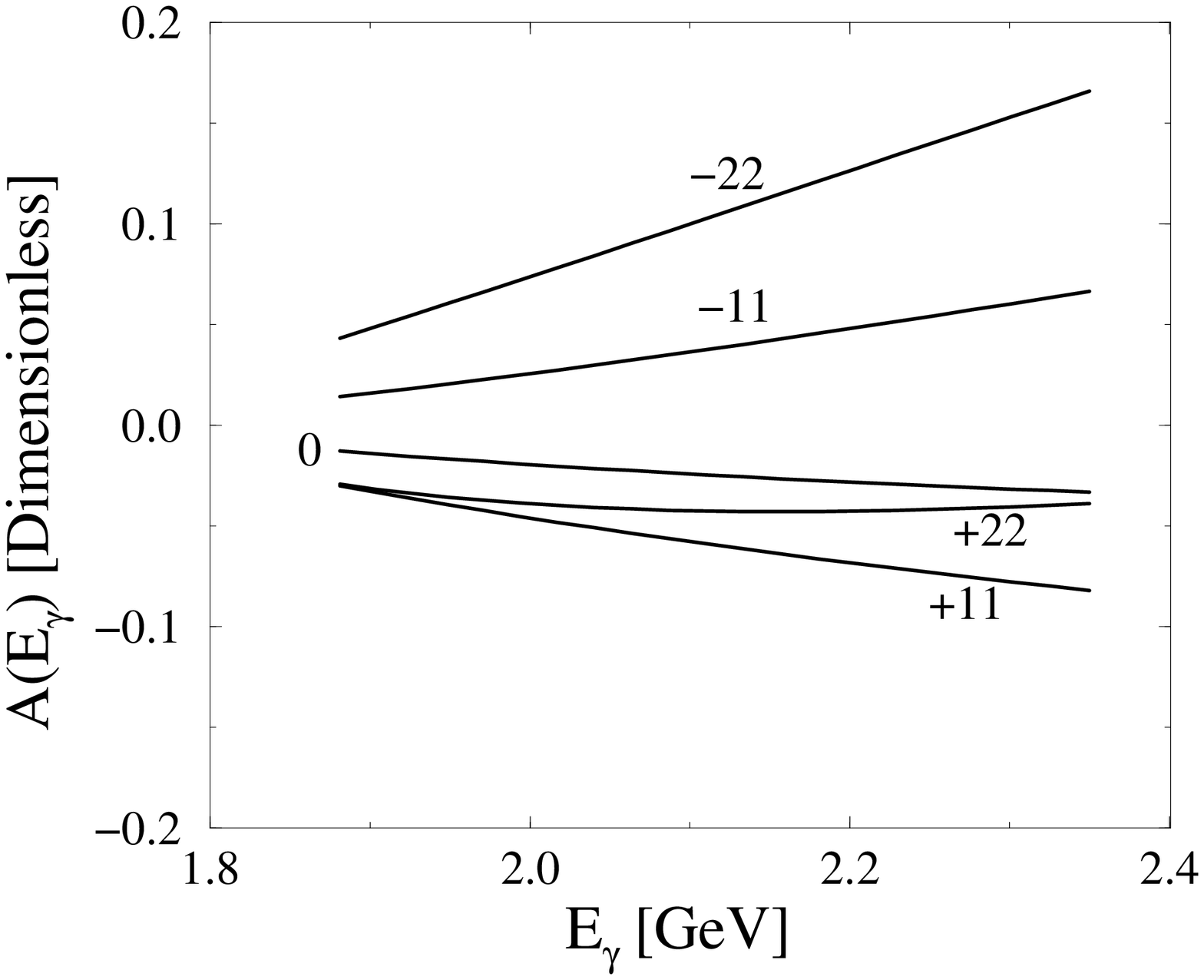}
\end{tabular}
\caption{Asymmetry $A(E_{\gamma},\theta)$ for the different photon
energies, $E_{\gamma}=1.8$ (upper-left), $2.0$ (upper-right)
and $2.2$ GeV (lower-left).  $A(E_{\gamma})$ of Eq.~(\ref{BATCS}) is
shown in the lower-right panel.  The labels on the figure denote the 
values of the $K^*N\Lambda^*$ coupling constant.}
\label{phi_dep2}
\end{figure}

In Fig.~\ref{phi_dep1}, $A(E_{\gamma},\theta)$ is depicted as a function
of $\theta$ for various terms separately at the photon energy 2.0 GeV.   
Since the $s$-, $u$-channels and contact term show weak
$\phi$-dependence for the whole range of $\theta$, the asymmetry turns
out to be almost zero, while the $K$-exchange asymmetry has
$A(E_{\gamma},\theta)=-1$ which follows from the fact that the cross
section vanishes at $\phi = 0$.  In contrast, $A(E_{\gamma},\theta)$
of $K^*$-exchange takes positive values except for $\theta =
0$ and $\pi$, which is due to the magnetic type coupling, though not
complete.  Therefore, we have established a characteristic difference
in three dominant terms; $A(E_{\gamma},\theta) \sim 0$ for the contact
term, $A(E_{\gamma},\theta) = -1$ for the $K$-exchange one and 
$A(E_{\gamma},\theta) \lesssim 0$ for the $K^*$-exchange one.
In Fig.~\ref{phi_dep2}, we show the asymmetries $A(E_{\gamma},\theta)$
for the proton target ($\gamma p \to K^+ \Lambda^*$) 
with all diagrams included as functions of $\theta$ and at different
photon energies (upper-left, right and lower-left panels).  Also shown 
is the $\theta$-integrated $A(E_{\gamma})$ as a function 
of $E_{\gamma}$ (lower-right panel).  We observe that in general the
asymmetry is small due to the presence of the contact term.  
As the strength of the $K^*N\Lambda^*$ coupling constant is increased,  
the asymmetry takes finite values either of positive (for the positive
$K^*N\Lambda^*$ coupling constant) or of negative value (for negative
$K^*N\Lambda^*$ coupling constant).   

Although it is experimentally more difficult to perform such a test in
the neutron reaction ($\gamma n \to K^0 \Lambda^*$), $K^*$-exchange
becomes important without the contact and $K$-exchange terms.  The
asymmetry, therefore, is expected to be positive in this case.    
\subsection{Polarized target}
If the target nucleon is polarized and at the same time if it is
possible to measure the polarization of $\Lambda^*$, we can check a
consistency for the spin of the produced particle ($\Lambda(1520)$)
owing to the conservation of its helicity.  Although it is probably a 
very difficult situation experimentally, we would like to see what may
be expected in observation.  
 
Suppose that we set the spin of the target proton to be $S_z(p)=+1/2$
and the photon polarization $S_z(\gamma)=+1$.  Then the total helicity
is 3/2.   In the forward angle scattering, the $z$-component of 
the total angular momentum is carried solely by the helicity.
Hence the forward cross section can take a finite value 
when the spin of $\Lambda(1520)$ is $S_z(\Lambda^*) = +3/2$. 
Otherwise, it vanishes.  This is a simple selection rule, but may be 
helpful for some reactions.  For instance, for the $\Lambda(1116)$-
production, since it has spin-1/2, the forward production under the
above condition, i.e. $S_z(p)=+1/2$ and $S_z(\gamma)=+1$ can not
occur.  As the scattering angle $\theta$ increases, however, the cross
section becomes finite.  

\begin{figure}[t]
\begin{tabular}{c}
\includegraphics[width=8cm]{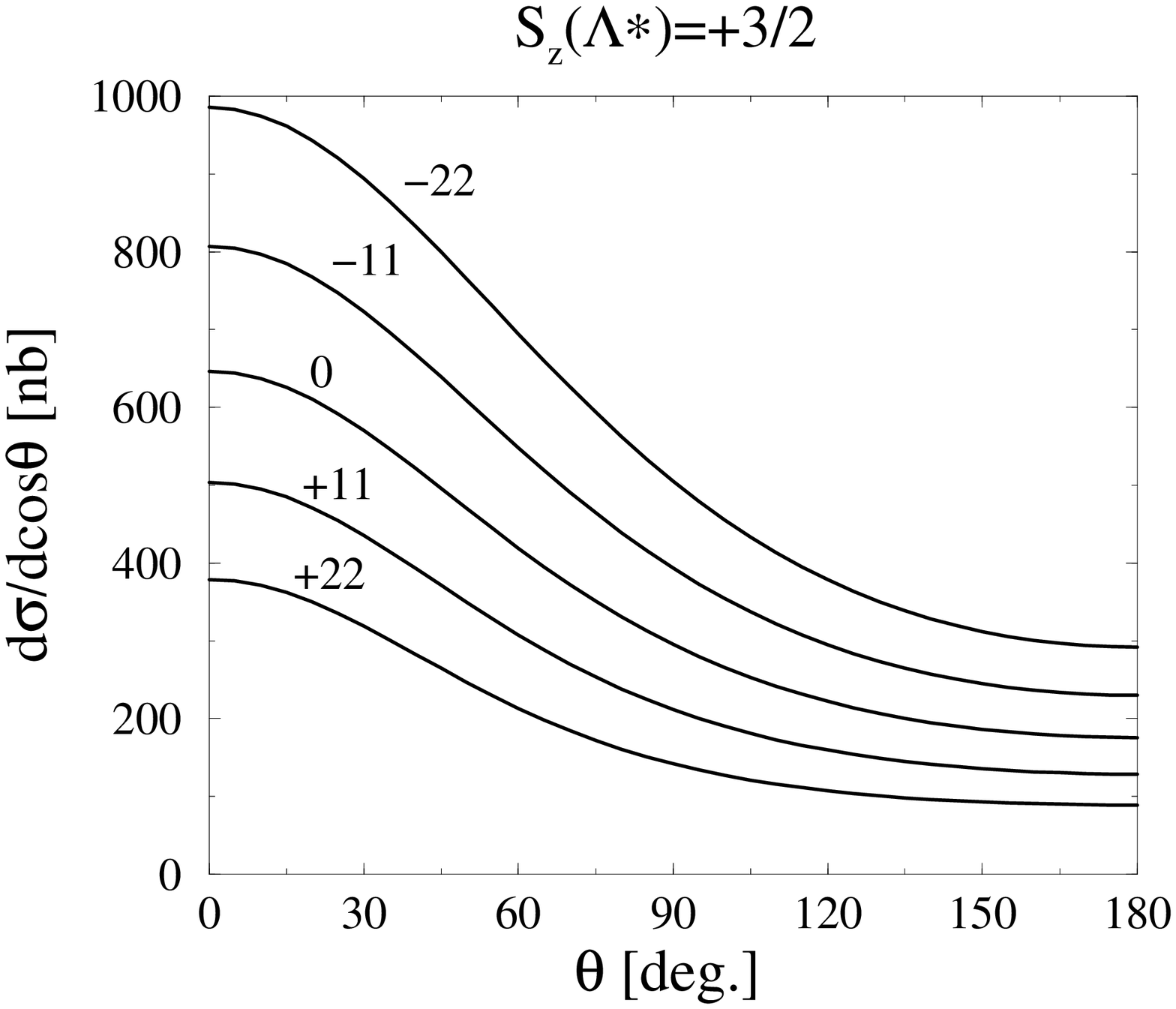}
\includegraphics[width=8cm]{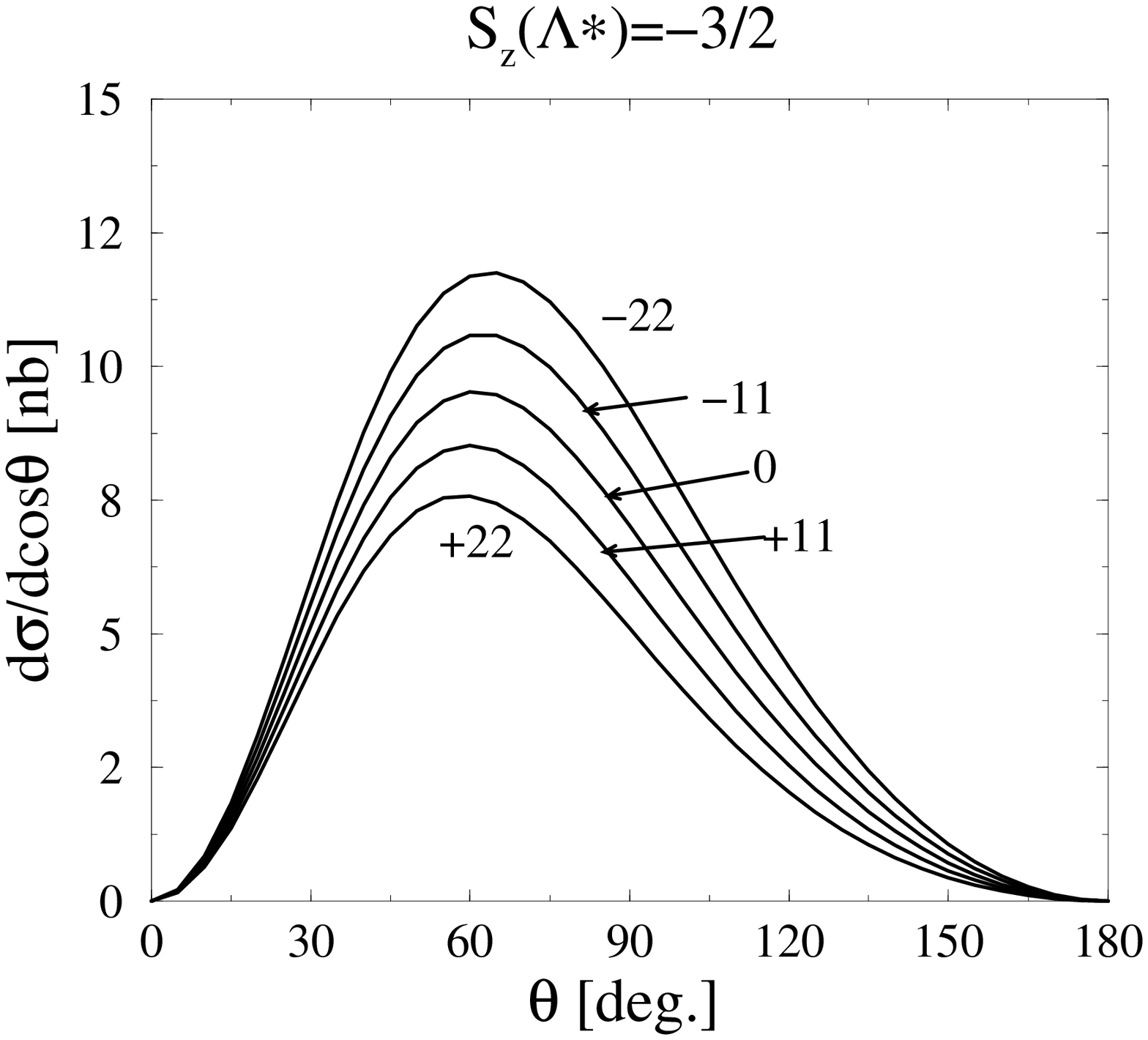}
\end{tabular}
\caption{The differential cross section of the reaction $\gamma p\to
K^+\Lambda^*$ with the polarized target at $E_{\gamma}=1.8$ GeV. The labels
on the figure denote the values of the $K^*N\Lambda^*$ coupling
constant.}     
\label{fig1}
\end{figure}

To see this situation explicitly, we show in Fig.~\ref{fig1} 
the differential cross sections for the polarized target $S_z(p)=+1/2$, 
the polarization of the final $\Lambda(1520)$ being 
either $S_z(\Lambda^*) = +3/2$ or $-3/2$.  The case of $S_z(\Lambda^*)
= +3/2$ is allowed while the other is suppressed, 
which is clearly seen in Fig~\ref{fig1}.  In this particular example,
it is interesting to see not only that the differential cross section
for $S_z(\Lambda^*) = -3/2$ becomes exactly zero at $\theta = 0$ and
$\pi$ but also that the magnitudes of these two cases are very much  
different: The suppressed differential cross section is very much
smaller than the allowed one by more than two order of magnitude.  
   
\subsection{Some implications to $\Theta^+$-photoproduction}
One of the motivations of the present work is to apply 
the reaction mechanism which can be established for the known resonance
$\Lambda(1520)$ to reactions involving $\Theta^+$.  
In fact, assuming that the spin and parity of $\Theta^+$ are the same as 
those of $\Lambda(1520)$, $J^P = 3/2^-$, we can apply the present 
results almost straightforwardly to the case of $\Theta^+$ within 
the framework of the effective Lagrangian approach.   
For example, we have already reported a large asymmetry between the
cross sections of the proton and neutron targets~\cite{Nam:2005jz}.  
The reaction rate of charge exchange process is significantly larger 
than that of charge non-exchange process.  
For the case of $\Lambda(1520)$ the proton target reaction is larger, 
while it is other way around for the case of $\Theta^+$, that is, the
neutron target reaction is dominant.    

In drawing this conclusion, however, it is crucial to know the strength 
of the $K^*N\Theta^+$ coupling constant, since the $K^*$-exchange  
process does not have a strong selectivity between the proton and
neutron target reactions.  As discussed in
Refs.~\cite{Nam:2005jz,Nam:2005jb}, the $K^*$-exchange process plays
an important role in angular distributions as well as in determining
the magnitude for the cross sections.  For the case of
$\Lambda(1520)$, a recent study indicates that the coupling is not
very strong and the above feature is maintained~\cite{Hyodo:2006uw}.    
On the contrary, the $K^*N\Theta^+$ coupling constant is not known,
but a naive estimation in the quark model is not so strong, and
therefore the above feature of charge exchange dominance is not very
much affected.   

We consider that the measurement and comparison of the experimental
and theoretical values of the $K^*N\Lambda^*$ coupling 
constant is important, in order to test the applicability of the
presently available models for hadrons, especially in the kaon
productions.  Once they are established, we can also extend the
same method to the physics of exotic hadrons including the $\Theta^+$, 
particularly because the energies of the $\Lambda(1520)$ and
$\Theta^+$ reactions are expected to be very similar. 
Such a strategy should be of great use in order to understand the
structure of exotic hadrons.  Once again, as shown in
Fig.~\ref{theta_dep}, the different $\phi$-dependence  
is helpful to determine the value for $g_{K^*N\Theta}$.  We would like
to emphasize also that the discussion of $\phi$-dependence is applied
to the resonance productions of spin-1/2.  

Another feature of the helicity conservation can be also applied to
the $\Theta^+$ productions.  Just as in the case of $\Lambda(1520)$, 
this will help us make some constraints on the unknown spin of the 
$\Theta^+$.   

\section{Summary and Conclusions}
In the present work, we have investigated the $\gamma N\to 
K^+\Lambda^*$ reaction in the effective Lagrangian approach.  In
particular, we aim at scrutinizing each contribution of various
diagrams to the reaction: On one hand, the contact term plays the
crucial role in describing the $\gamma p\to K^+\Lambda^*$ process.
On the other hand, the $K^*$-exchange term may become non-negligible.
It is of utmost importance to understand this difference, since the
contact term exists in the proton channel while it is absent in the
neutron channel, so that the rate of the $\Lambda^*$-production turns
out to be suppressed in the neutron target.  

We have asserted that it is of great use to investigate the azimuthal
angular distributions of the differential cross section in order to
understand the role of each contribution.  In particular, the angular 
$\phi$-distribution, or alternatively, the asymmetries
$A(E_{\gamma},\theta)$ and $A(E_{\gamma})$, which can be measured by 
using the polarized photon beam, distinguish each role of the contact
term, $K^*$-exchange, and $K$-exchange terms according to the
polarization of the photon beam.  We have observed that if the contact
term dominates, the asymmetry becomes almost zero, while the
$K^*$-exchange contribution increases, the asymmetry becomes finite.
Hence, a measurement of the asymmetries will provide a touchstone for 
understanding the reaction mechanism of the $\Lambda^*$-photoproduction. 

Not only theoretically but also experimentally, the contact term
dominance is one of the most interesting features for the 
spin-3/2 baryon photoproduction as well as for the
$\Lambda^*$-photoproduction in general.  In spite of possible model 
dependence in treating baryons with different spins, the
above-mentioned feature does not appear in the spin-1/2 baryon
photoproduction.  An experimental confirmation of the contact term
dominance in the $\Lambda^*$-photoproduction is important and may be a
challenging task.  As discussed in the present work, since the test of
the contact term dominance can be made in particular in the very
forward direction and near the threshold, the LEPS collaboration at
the SPring-8 may be one of the good places for verifying it. 
\section*{Acknowledgments}
The present work is supported by the Korea Research Foundation Grant
funded by the Korean Government(MOEHRD) (KRF-2006-312-C00507).  
We are grateful to T.~Nakano and A.~Titov for fruitful discussions. 
S.i.N. would like to thank K.~Horie, S.~Shimizu, J.~K.~Ahn, T.~Hyodo and
T.~Sawada for interesting and stimulating discussions about this  
work. S.i.N. also  appreciates the hospitality during his stay at RCNP
in Japan, where part of this work was performed. The work of A.H. is 
supported in part by the Grant for Scientific Research ((C)
No.16540252) from the Education, Culture, Science and Technology of 
Japan.  The work of S.i.N. is supported in part by the Brain Korea 21 
(BK21) project in Center of Excellency for Developing Physics
Researchers of Pusan National University, Korea. 

 
\end{document}